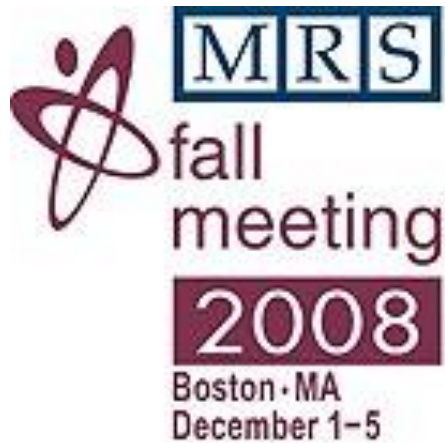

# Optical Characterization of Nanocrystallized AlN:Er Films Prepared by Magnetron Sputtering

**V. Brien[1\*],** M. Dossot[2], H. Rinnert[3], S. S. Hussain[1], B. Humbert[2] and P. Pigeat[1]
[1]CNRS/Nancy-Université (LPMIA UMR CNRS 7040), Boulevard des Aiguillettes, B.P. 239, F-54506 Vandœuvre-lès-Nancy, France
[2]CNRS/Nancy-Université (LCPME UMR 7564 CNRS), 405, rue de Vandoeuvre, F-54600 Villers-lès-Nancy, France
[3]Nancy-Université/CNRS (LPM UMR CNRS 7556), Boulevard des Aiguillettes, B.P. 239, F-54506 Vandœuvre-lès-Nancy, France

\*Corresponding author

ABSTRACT

Four groups of AlN:Er films were fabricated by r.f. reactive magnetron sputtering on silicon substrates: nano-columns, feather-shaped nanostructures, nano-rods and equiaxed nano-grains. High spectral resolution visible photoluminescence measurements were performed at room temperature. The photoluminescence spectrum of the nano-columnar films shows a broad emission band centered around 530 nm and a superimposed fine structure attributed to nano-AlN photoluminescence and to $^4S_{3/2} \to {}^4I_{15/2}$ and $^2H_{11/2} \to {}^4I_{15/2}$ erbium transitions, respectively. The fine structure of erbium emission is shown to depend on the matrix nano-structure. The more crystallized films (nano-columnar) produce the more intense and finest erbium peaks. The equiaxed nanograins favor the $^2H_{11/2}$–$^4I_{15/2}$ contribution to the detriment of the $^4S_{3/2}$–$^4I_{15/2}$ one.

# INTRODUCTION

In the field of optoelectronic technologies, the need for integration of future devices forces to seek materials emitting high luminescence in the visible or near-infrared spectral ranges. Rare earth ions inserted in host matrix are of potential interest for their high luminescence properties. In recent years, the increasing demand of sources and optical amplifiers working at wavelengths compatible with the telecommunication technologies based on optical fibers renewed the interest in works dealing with luminescence of rare earths cations, notably $Er^{3+}$ [1]. The interest for this cation with a $4f^{11}$ electronic configuration comes from the electronic transitions between the two states $^4I_{13/2}$ and $^4I_{15/2}$ likely to produce luminescence at 1.54 µm, which coincides with the low-loss region of silica-based optical fibers. Due to selection rules, the parity forbidden $^4I_{13/2} \rightarrow {^4I_{15/2}}$ transition of isolated erbium ions induces a weak absorption cross-section. Several teams have developed new solutions for some years that are based on the insertion of erbium cations within a material having a higher absorption cross-section, and benefiting from indirect excitation processes [2-11]. The nature and structure of the host matrix for the luminescent ions are crucial parameters to be selected in view of the final properties wanted for the optoelectronic device. Among III-V components, aluminum nitride (AlN) semiconductor is interesting thanks to its wide gap (6.2 eV) that has been shown to be useful to reduce the temperature quenching effect induced by free carriers [12-13]. Despite some theoretical works devoted to a structure-related $Er^{3+}$ excitation / emission mechanism, the photoluminescence (PL) of rare earth cations inserted in wide gap matrices is still not well understood [14-15]. It is thought that energy transfer between the host matrix and rare earth cations strongly increases the luminescence intensity. In this case, the electronic levels brought by the matrix defects could play an important role in promoting this transfer. By contrast, defects can also be responsible for non-radiative recombination of excitons, limiting the photoluminescence intensity. Preliminary works on AlN:Er systems show anyhow that the matrix structure is a relevant parameter to control the photoluminescence properties [16].

The main aim of the present work is to make reference samples with various morphologies of the AlN host matrix to study the effect of the nano-structuration of the matrix on the erbium visible luminescence properties. To achieve this goal, room temperature visible photoluminescence of AlN films deposited by r.f. reactive magnetron sputtering and doped with erbium cations will be measured for samples with variable nano-morphology (columnar, granular with equiaxed, rod-shaped or feather-shaped grains).

# EXPERIMENT

The radio frequency (r.f.) magnetron sputtering technique was chosen to deposit the samples because of its known capacity to be an industrial low-cost fabrication process. In this study, AlN:Er thin films were sputtered on [001] mono-crystalline silicon substrates at room temperature. The sputtering system was pumped down to a residual ultra high vacuum pressure of $10^{-6}$ Pa of $H_2O$ controlled by mass spectroscopy. Sputtering was performedin a mixed $Ar/N_2$ atmosphere (50/50) under a total sputtering pressure P. The gas purities were 99.999 %. The insertion of erbium in AlN samples was obtained by co-sputtering the two Al and Er elements thanks to the use of an aluminum target disk (purity of 99.9%), whose one sector was replaced by Er bulk material (purity of 99.99 %). The diameter of the target was 60 mm and the target-sample distance was 150 mm. The target was systematically sputter cleaned for 15 min using an Ar plasma prior to any deposition. The r.f. power delivered by the 13.56 MHz generator (W) was varied between 50 and 300 W (see Table I). The bias voltage (U) was also varied between 0 and 100 V. The different nano-structures were prepared thanks to adequately chosen different sputtering parameters that are compiled in Table I.

**Table I:** Magnetron sputtering parameters used to elaborate the nano-structured Er:AlN films.

| Nano-structure | W (W) | P (Pa) | U (V) | Common parameters |
|---|---|---|---|---|
| A: Columns | 300 | 0.5 | 0 | T = Room temperature |
| B: Rods | 300 | 0.5 | 50 | Target –substrate distance = 150 mm |
| C: Feathers | 50 | 0.4 | 0 | $Ar/N_2$ = 50/50, Total gas volume = 5 sccm |
| D: Equiaxed | 300 | 0.5 | 100 | Residual pressure = $1.10^{-6}$ Pa of $H_2O$ |

More details on process can be found in previous works performed on growth mechanisms of AlN films [17-20]. The thickness of the AlN films was controlled in real-time during the growth of the layer thanks to an interferential optical reflectometer and was confirmed by transmission electron microscopy (TEM) cross-section observations. The chemical composition of the samples was determined by Electron Dispersion of X-rays and was found to be within the 0.1 – 1 atomic % range. Due to the low erbium concentration, the insertion of erbium did not change the growth modes observed on pure AlN films, as demonstrated by the TEM investigation.

TEM observations were performed on a PHILIPS CM20 microscope operating at an accelerating voltage of 200 kV using the technique of microcleavage of samples. Visible room-temperature photoluminescence was measured with high spectral resolution (HR-PL) using a Jobin-Yvon T64000 spectrometer equipped with a 1800 gr/mm grating and a liquid $N_2$ cooled CCD detector. The exciting laser wavelength was either the 458 nm or the 514 nm emission lines of an $Ar^+$ laser (Stabilite 2017 from Spectraphysics). The samples were mounted on a confocal microscope stage (pinhole diameter 100 μm) and the photoluminescence was collected by a ×50 microscope objective with Numerical Aperture = 0.55. The photoluminescence light was separated from the incident laser light using proper Notch filters.

**RESULTS AND DISCUSSION**

After characterization by TEM, AlN:Er films morphologies could be categorized in four groups: granular with nano-columns (A), granular with rod-shaped crystallites (B), granular with grains feather-shaped crystallites (C) and granular with equiaxed crystallites (D). The average diameter of the columns was found to range from 10 to 25 nm, depending on the sample. The average width of the rods was found to be equal to 11 nm. The average width of the rods that composed the feather-shaped grains was found to range from 6 to 24 nm and the average diameter of the equiaxed grains was found to range from 3 to 20 nm, depending on the sample. More details on the morphological characterization could be found in the previously published works [17-20].

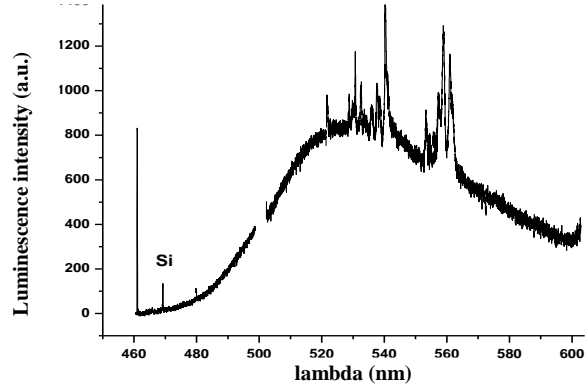

**Figure 1:** High spectral resolution photoluminescence spectrum obtained at room temperature on the AlN:Er columnar film with the incident light λ = 458 nm in the visible domain.

Figure 1 shows the typical visible HR-PL spectrum recorded at room temperature in the [460 nm – 600 nm] range of an erbium-doped nano-columnar film (type A) excited at 458 nm. The signal is composed of a wide emission band ranging from 2.6 eV (480 nm) to 2.1 eV (600 nm) centered at 2.3 eV (530 nm), and of two distinct groups of fine peaks localized at [528 nm – 545 nm] and [545 nm – 570 nm], respectively. Such a broad band (480 nm – 600nm) has already been recorded in PL spectra of many different types of aluminum nitride nano-objects (nanopowders, nanocrystalline layers, nanowhiskers, nanowires…) [21-24]. The AlN photoluminescence is by itself an interesting subject studied in several works dealing with photonic applications. In spite of the fact that theoretical interpretation of this broad PL emission is not unanimous, all the interpretations consider that several defects can contribute to the spectral shape of the PL emission of nano-structured AlN samples. These defects are located at different energy values attributed by the authors to either nitrogen vacancies, oxygen point defects substituting nitrogen atoms or interstitial aluminum atoms.

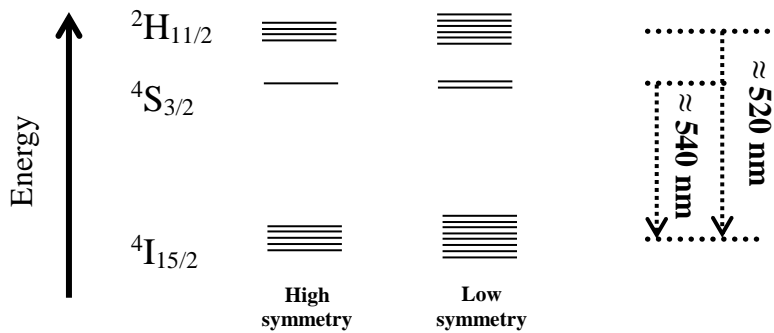

Figure 2: Energy levels or $Er^{3+}$ and observed radiative transitions

Figure 1 also reveals that two groups of sharp peaks are superimposed to the broad PL band of the AlN matrix. These peaks can be attributed to radiative electronic transitions of doping erbium cations. The first group (528 nm to 545 nm) and the second one (545 nm to 570 nm) can be respectively assigned to $^2H_{11/2} \rightarrow {}^4I_{15/2}$ (≈520 nm) and to $^4S_{3/2} \rightarrow {}^4I_{15/2}$ (≈540 nm) electronic transitions (cf. Figure 2). The obtained experimental $^2H_{11/2}$–$^4I_{15/2}$ bunch and the $^4S_{3/2}$– experimentally; 2) Erbium cations are localized on at least two kinds of doping sites with high symmetry (for instance, octahedral or tetragonal sites) and the spectrum is the superimposition of all these contributions.

$^4I_{15/2}$ bunch can be respectively decomposed in at least 10 and 6 spectral contributions. Due to the properties of $Er^{3+}$ PL and selection rules, this decomposition can be interpreted according to two hypothesizes: 1) erbium cations are all localized in a same kind of low-symmetry site, which would induce a strong splitting of the degeneracy of the electronic levels. In that case, the number of peaks detected in the two groups is not enough to account for all possible transitions (6 x 8 = 48, 2 x 8 = 16 peaks would theoretically be expected for the $^2H_{11/2}$–$^4I_{15/2}$ bunch and the $^4S_{3/2}$–$^4I_{15/2}$ bunch, respectively) but spectral overlap may prevent to resolve all peaks

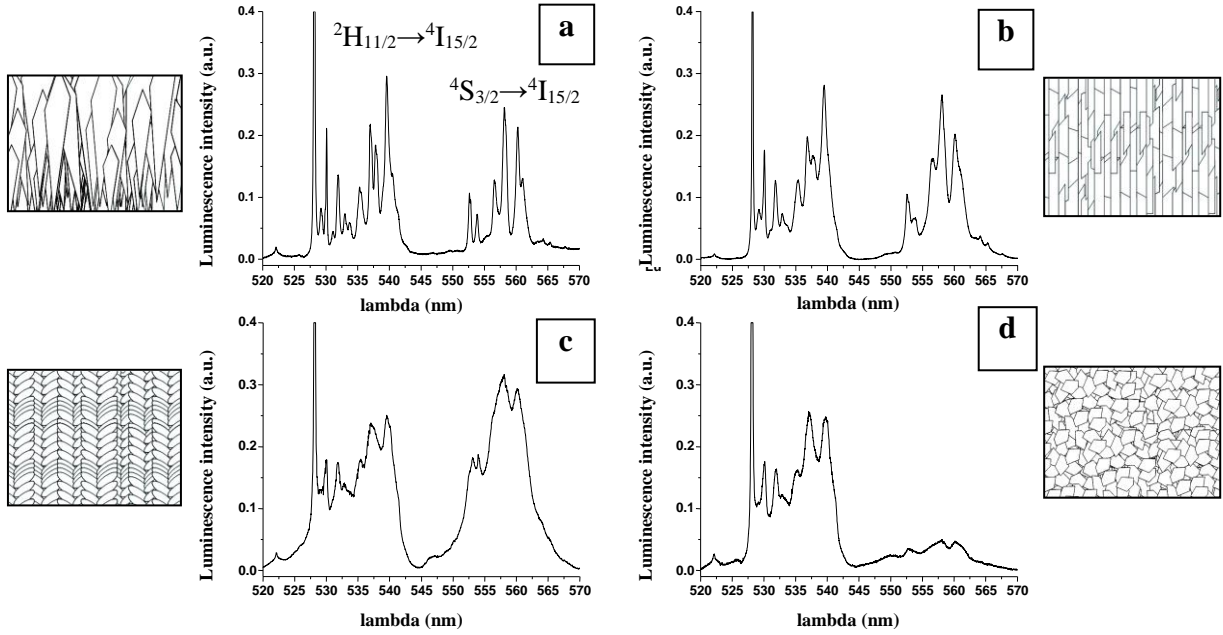

Figure 3: Photoluminescence spectra obtained at room temperature on the AlN:Er films in the 520 – 570 nm domain with the incident light of 514 nm: (a) Type A: nano columnar 16 nm (b) Type B: nano rods 20 nm (c) Type C: nano feathers 6-20 nm(d) Type D: equiaxed nano granular 3 nm.

In order to investigate the role of the structure on the PL properties of $Er^{3+}$ in AlN, the HR-PL was performed for the different morphologies and is presented in figure 3. All spectra were normalized in intensity to the Si substrate phonon mode at 520 cm$^{-1}$. The broad PL band was subtracted for the clarity of the presentation. The spectra show that there are some subtle changes concerning the fine structure of the two bunches of peaks and a more obvious change concerning the integrated emission intensity ratio of $^4S_{3/2}$–$^4I_{15/2}$ group over the $^2H_{11/2}$–$^4I_{15/2}$ group. This ratio appears to depend on the AlN:Er film structure. Table II collects the values of this ratio and it is clear that these values seem to be correlated to the crystallization degree of the AlN matrix. This degree may allow changing the amount of non-radiative process, via electronic states created by the modification of the dangling bonds and/or vacancies density because of a change in the size and shape of grains.

Indeed, this ratio strongly decreases for the structure D that is the less crystallized AlN film, whereas the feather-shaped grains (structure C) correspond to the higher value. The change of the emission intensity certainly results from the variation of the Stark effect that splits the degeneracy of the electronic levels and relaxes the selection rules. The matrix nano-structuration may then be probed by $Er^{3+}$ cations, and the modulation of the $^4S_{3/2} \rightarrow {^4I_{15/2}}$ / $^2H_{11/2} - {^4I_{15/2}}$ intensity ratio could be very interesting for photonic applications.

Table II: Detail of the calculation of the ratio of the total integrated intensities of the $^4S_{3/2} - {^4I_{15/2}}$ peak group over the $^2H_{11/2} - {^4I_{15/2}}$ peak group as a function of the AlN:Er film structure.

| Sample | $^2H_{11/2} - {^4I_{15/2}}$ [528-545 nm] | $^4S_{3/2} - {^4I_{15/2}}$ [545-570 nm] | Ratio $^4S_{3/2}$ / $^2H_{11/2}$ |
|---|---|---|---|
| A: Nano columns 16 nm | 1.28 | 1.09 | 0.85 |
| B: Nano rods 11 nm | 1.41 | 1.37 | 0.97 |
| C: Nano feathers 6 - 20 nm | 2.22 | 2.90 | 1.31 |
| D: Nano equiaxed grains 3 nm | 2.07 | 0.56 | 0.27 |

As the Stark effect probably plays an important role on the Er-related luminescence, the study of the location of the erbium sites is necessary. A precise understanding of the kinds of sites occupied by erbium cations requires further experiments such as site-selective excitation by changing the excitation wavelength and low temperature HR-PL measurement to decrease the peak width and then to increase the spectral resolution and the accuracy of spectral decomposition. Besides, time-resolved PL measurements would enable to determine exciton dynamics and then would bring about interesting information concerning the PL mechanism.

## CONCLUSIONS

Er-related and defect-related luminescence in the visible range was obtained in Er-doped AlN thin films prepared by magnetron sputtering. The broad PL band around 550 nm is attributed to defects of the AlN matrix. The fine spectral features of $Er^{3+}$ centers, attributed to $^4S_{3/2} \rightarrow {^4I_{15/2}}$ and $^2H_{11/2} \rightarrow {^4I_{15/2}}$ erbium transitions, was found to be strongly correlated to the different nano-structurations of the AlN matrix explored in the present work. This result is attributed to the Stark effect, which is dependent on the structural environment of $Er^{3+}$ ions.

## REFERENCES

1. A.J. Kenyon, *Progress in Quantum electronics*, **26**, 225-284 (2002)
2. A. R. Zanatta, C. T. M. Ribeiro and U. Jahn*, J. Appl. Phys*. **98** 1-8 (2005)
3. J. W. Lim, W. Takayama, Y. F. Zhu, J. W. Bae, J. F. Wang, S. Y. Ji, K. Mimura, J. H. Yoo and M. Isshiki, *Mater. Lett*., **61**, 17,3740-5 (2007)
4. J.C. Oliveria, A. Cavaleiro, M.T. Vieira, L. Bigot, C. Garapon, J. Mugnier and B. Jacquier, *Thin Solid Films* **446** 264 – 270 (2004)
5. V. I. Dimitrova, P. G. Van Patten, H. Richardson, M. E. Kordesch, *Appl. Surf. Sci.* **175-176**, 480-483 (2001)
6. X. Wu, U. Hömmerich, J. D. MacKenzie, C. R. Abernathy, S. J. Pearton, R.G. Wilson, R. N. Schwartz and J. M. Zavada, *Journal of Luminescence* **72-74** 284-286 (1997)

7. S. B. Aldabergenova, M. Albrecht, H. P. Strunk, J. Viner, P. C. Taylor, A. A . Andreev, *Mat. Sci. Eng. B* **81**, 144 – 146 (2001)
8. F. Priolo, G. Franzo, S. Coffa, and A. Carnera, *Phys. Rev. B* **57**, 4443 (1998)
9. M. Fujii, M. Yoshida, Y. Kansawa, S. Hayaski, and K. Yamamoto, *Appl. Phys. Lett*. **71**, 1198, (1997)
10. P. G. Kik, M.L. Brongersma and A. Polman, *Appl. Phys. Lett*. **76**, 2325 (2000)
11. G. Wora Adeola, O. Jambois, P. Miska, H. Rinnert, and M. Vergnat, *Appl. Phys. Lett*. **89**, 101920 (2006)
12. P.N. Favennec, H. L'Haridon, D. Moutonnet and Y.L. Guillo. *Electron. Lett.* **25** (1989), p. 718
13. S.J. Pearton, C.R. Abernathy, J.D. MacKenzie, U. Hommerich, X. Wu, R.G. Wilson, R.N. Schwartz, J.M. Zavada and F. Ren. *Appl. Phys. Lett.* **71** (1997), p. 1807
14. S.B. Aldabergenova, M. Albrecht, A.A. Andrrev, C. Inglefield, J. Viner, V.Y. Davydov, P.C. Taylor and H.P. Strunk, *J. Non-Cryst. Solids* **283** (2001), p. 173.
15. J.-W. Lim, W. Takayama, Y.F. Zhu, J.W. Bae, J.F. Wang, S.Y. Ji, K. Mimura, J.H. Yoo, M. Isshiki, *Curr. Appl. Phys.*, **7**, 3 (2007) 236-239
16. V. Brien, P. Miska, H. Rinnert, D. Genève, P.Pigeat, *Mater. Sci. and Eng. B*, **146** (1-3) (2008) 200-203
17. V. Brien and P. Pigeat, *J. Cryst. Growth* **299**, 189 (2007)
18. V. Brien and P. Pigeat, *J. Cryst. Growth,* **310**, 16, 3890-3895 (2008),
19. V. Brien, P. Miska, B. Bolle, P. Pigeat, *J. Cryst. growth*, **307/1**, 245-252 (2007)
20. P. Pigeat, T. Easwarakhanthan, *Thin Solid Films*, **516**, 12, 3957 (2008)
21. Y. C. Lan, X. L. Chen, Y. G. Cao, Y. P. Xu, L. D. Xun, , T. Xu, J. K. Liang, *J. Cryst. Growth*, **207** 247-250 (1999)
22. J. Siwiec, A. Sokolowska, A. Olszyna, R. Dwilinski, M. Kaminska, J. Konwerska-Hrabowska, *Nanostruct. Mater*., **10**, 4, 625-634 (1998)
23. H.T. Chen, X.L. Wu, X. Xiong, W.C. Zhang, L.L. Wu and J. Zhu et al., *J. Phys. D: Appl. Phys.* **41** 025101 (5pp) (2008)
24. C. Xu, L. Yin, G. Wang, *Phys. Sta. Sol*. (a) **2** 329-335 (2003)